\documentclass[12pt]{iopart}
\usepackage{hyperref}

\hypersetup{
 	colorlinks=true,
 	linkcolor=red,
 	citecolor=blue,
 }

\begin{document}
\title{On the linear transformation between inertial frames}

\author{Qing Gao}
\address{School of Physical Science and Technology, Southwest University, Chongqing 400715, China}
\ead{gaoqing1024@swu.edu.cn}

\author{Yungui Gong}
\address{School of Physics, Huazhong University of Science and Technology, Wuhan, Hubei 430074, China}
\ead{Corresponding author. yggong@hust.edu.cn}

\begin{abstract}
In the derivation of Lorentz transformation,
linear transformation between inertial frames is one of the most important steps.
In teaching special relativity,
we usually use the homogeneity and isotropy of spacetime to argue that the transformation must be linear transformation without providing any rigorous detail.
Here in the first time we provide a solid mathematical proof of the argument that the transformation between two inertial frames must be linear because of the homogeneity and isotropy of spacetime.
\end{abstract}


\maketitle

\section{Introduction}
In 1905 Einstein proposed special relativity based on two postulates: postulate of relativity and postulate of the constancy of the speed of light.
The postulate (principle) of relativity states that the laws of physics have the same
form with respect to all inertial systems.
The postulate of the constancy of the speed of light means that the speed of light is a finite
constant $c$,
independent of the motion of its source and observers.
Based on these two postulates, we can derive Lorentz transformation and the invariance of spacetime interval.
In the derivation, linear transformation between inertia systems is one of the essential ingredients.
For alternative derivation of Lorentz transformation, see references \cite{macdonald,levy,young,heras,rivas}.
In the literature and many textbooks, linear transformation was treated as either an assumption or a well known fact without any rigorous proof, see for example references \cite{heras,rivas,ref01,ref0,ref03,jerrold,ref2}.

In \cite{ref2}, the argument of linear transformation is
that the transformation equations are linear because a nonlinear transformation could yield an acceleration in one system even if
the velocity were constant in the other,
but Kleppner and Kolenkow didn't provide any further detail about it.
The same argument was presented in Rindler's book \cite{ref3} and he gave a proof as follows. Consider a standard clock C freely moving
through $S$, its motion being given by $x_i=x_i(t)$, where $x_i$ ($i=1,2,3$) stands for $(x,y,z)$. Then $dx_i/dt=\rm{constant}$.
If $\tau$ is the time indicated by C itself,
homogeneity requires the constancy of $dt/d\tau$.
(Equal outcomes here and there, now and later,
of the experiment that consists of timing the ticks of a standard clock moving at constant speed.) Together these results
imply $dx_\mu/d\tau=\rm{constant}$ and thus $d^2x_\mu/d\tau^2=0$, where we have written $x_\mu$ ($\mu=1,2,3,4$) for $(x,y,z,t)$.
In $S'$ the same argument yields $d^2x'_\mu/d\tau^2=0$. But we have
\begin{eqnarray}
\frac{dx'_\mu}{d\tau}=\sum\frac{\partial x'_\mu}{\partial x_\nu}\frac{dx_\nu}{d\tau}, \nonumber\\
\frac{d^2x'_\mu}{d\tau^2}=\sum\frac{\partial x'_\mu}{\partial x_\nu}\frac{d^2x_\nu}{d\tau^2}
+\sum\frac{\partial^2 x'_\mu}{\partial x_\nu\partial x_\sigma}\frac{dx_\nu}{d\tau}\frac{dx_\sigma}{d\tau}.\nonumber
\end{eqnarray}
Thus for any free motion of such a clock the last term in the above line of equations must vanish. This can only happen if
$\partial^2x'_\mu/\partial x_\nu x_\sigma=0$; that is, if the transformation is linear.

In the above proof, the constancy of $dt/d\tau$ and the same argument applied in $S'$  implicitly assume the invariance of the proper time (or the spacetime interval) $d\tau$. Actually Weinberg gave the proof that a general coordinate transformation that leaves the invariant the proper time must be linear transformation in his book \cite{ref4}.
The proof is as follows. A general coordinate transformation $x\to x'$ will change $d\tau$ into $d\tau'$, given by
\begin{eqnarray}
\label{s2}
d\tau^{\prime 2}=-\eta_{\alpha\beta}dx^{\prime \alpha} dx^{\prime \beta}=-\eta_{\alpha\beta}\frac{\partial x^{\prime \alpha}}{\partial x^\mu}\frac{\partial x^{\prime \beta}}{\partial x^\nu}dx^\mu dx^\nu.\nonumber
\end{eqnarray}
If this is equal to $d\tau^2=-\eta_{\mu\nu}dx^\mu dx^\nu$ for all $dx^\mu$, we must have
\begin{eqnarray}
\label{eta}
\eta_{\mu\nu}=\eta_{\alpha\beta}\frac{\partial x^{\prime\alpha}}{\partial x^\mu}\frac{\partial x^{\prime\beta}}{\partial x^\nu}.\nonumber
\end{eqnarray}
Differentiation with respect to $x^{\gamma}$ gives
\begin{eqnarray}
\label{eta1}
0=\eta_{\alpha\beta}\frac{\partial^2 x^{\prime\alpha}}{\partial x^\gamma \partial x^\mu}\frac{\partial x^{\prime\beta}}{\partial x^\nu}+\eta_{\alpha\beta}\frac{\partial x^{\prime\alpha}}{\partial x^\mu}\frac{\partial^2 x^{\prime\beta}}{\partial x^\nu \partial x^\gamma}.\nonumber
\end{eqnarray}
To solve for the second derivatives, we add to this the same equation with the interchange $\gamma\leftrightarrow \mu$,
and substract the same with the interchange $\gamma \leftrightarrow \nu$;
then we are left with
\begin{eqnarray}
\label{eta2}
0=2\eta_{\alpha\beta}\frac{\partial^2 x^{\prime\alpha}}{\partial x^\gamma \partial x^\mu}\frac{\partial x^{\prime\beta}}{\partial x^\nu}.\nonumber
\end{eqnarray}
But both $\eta_{\alpha\beta}$ and $\partial x^{\prime \beta}/\partial x^\nu$ are nonsingular matrices, so this immediately yields
\begin{eqnarray}
\label{eta3}
\frac{\partial^2 x^{\prime\alpha}}{\partial x^\mu\partial x^\nu}=0.\nonumber
\end{eqnarray}
The general solution is of course a linear function, therefore the linear transformation is proved. This proof assumes the invariance of the proper time $d\tau$.
Weinber further pointed out that the transformations $x\to x'$ that leave $d\tau$ invariant when $d\tau=0$ are in general nonlinear which form the conformal group \cite{ref4}.
Therefore, it is necessary to provide a pedagogical method to help students understand the argument that homogeneity and isotropy of spactime means that the coordinate transformations between inertia frames must be linear transformations.
After all, the science of physics is about explaining physical idea with mathematical formula.
In this paper, we use the two postulates and the assumption of the homogeneity and isotropy of spacetime to prove that
the general coordinate transformations between inertial frames must be linear transformations.

\section{Proof of linear transformation}

Consider two inertial frames $\Sigma$ and $\Sigma'$ with $\Sigma'$ moving with respect to $\Sigma$ at a constant speed $v$ ($v\neq c$) along the $x$ axis.
We suppose that the coordinate and time in each inertial frame are defined based on standard method.
Initially, the clock at the origin of $\Sigma'$ was synchronized with the clock at the origin of $\Sigma$, i.e, $x=0$, $t=0$, $x'=0$, $t'=0$ (in principle any point can be chosen to synchronize the clocks, for convenience we choose the origins of the coordinates).
A general transformation between $\Sigma'$ and $\Sigma$ is
\numparts
\begin{eqnarray}
\label{eq1a}
x'=f(x,t),\\
\label{eq1b}
t'=g(x,t),
\end{eqnarray}
\endnumparts
and the differential forms are
\numparts
\label{eq2}
\begin{eqnarray}
\label{eq2a}
dx'=\frac{\partial f}{\partial x}dx+\frac{\partial f}{\partial t}dt,\\
\label{eq2b}
dt'=\frac{\partial g}{\partial x}dx+\frac{\partial g}{\partial t}dt,
\end{eqnarray}
\endnumparts
where the functions $f$ and $g$ are arbitrary functions of two variables.

{\bf Lemma 1}: The function $f$ is a function of a single variable, $f(x,t)=f(x-vt)$ with $f(0)=0$.

The origin $x'=0$ of $\Sigma'$ (it could be any point) moves at a constant speed $v$ with respect to $\Sigma$, the motion in $\Sigma$ is $dx=vdt$ and its motion in $\Sigma'$ is $dx'=0$.
Substituting $dx=vdt$
into Eq. \eref{eq2a}, we get
\begin{eqnarray}
\label{eq3}
dx'=(\frac{\partial f}{\partial x}v+\frac{\partial f}{\partial t})dt=0,
\end{eqnarray}
so
\begin{eqnarray}
\label{ft}
\frac{\partial f}{\partial x}=-\frac{1}{v}\frac{\partial f}{\partial t}.
\end{eqnarray}
Take the time derivative on both sides of Eq. \eref{ft}, we get
\begin{eqnarray}
\label{ft1}
\frac{\partial^2 f}{\partial x\partial t}=-\frac{1}{v}\frac{\partial^2 f}{\partial t^2}.
\end{eqnarray}
Therefore, the function $f$ should be a linear function if $\partial^2 f/\partial x\partial t=0$.
Taking the spatial derivative on on both sides of Eq. \eref{ft} and combining the result with Eq. \eref{ft1}, we get
\begin{eqnarray}
\label{eq4}
\frac{\partial^2 f}{\partial x^2}-\frac{1}{v^2}\frac{\partial^2 f}{\partial t^2}=0.
\end{eqnarray}
It is obvious that the function $f$ satisfies the wave equation, so the solution is
\begin{eqnarray}
\label{f}
f=f(x-vt).
\end{eqnarray}
Therefore, here we prove that the function $f(x,t)$ is an arbitrary function $f(p)$ of a single variable $p$.

According to the principle of relativity, the inverse transformation is $x=f(x'+vt')$.
In other word, $x'+vt'=f+vg=f^{-1}(x)$ is a function of $x$ only and independent of $t$, so the time derivative is zero and we have
\begin{equation}
\label{fgrel2}
\frac{\partial f}{\partial t} =-v \frac{\partial g}{\partial t},
\end{equation}
and
\begin{equation}
\label{fgrel3}
\frac{\partial f}{\partial x} = \frac{\partial g}{\partial t}.
\end{equation}

{\bf Lemma 2}: The function $g$ is a function a single variable, $g(x,t)=g(x-\tilde{u} t)$ with $g(0)=0$, where $\tilde{u}$ is an unknown constant independent of the spacetime coordinate.

Consider a body moving at a constant speed along $x$ axis,
the motion in $\Sigma$ is $dx=udt$ and the motion in $\Sigma'$ is $dx'=u'dt'$.
Combining Eqs. \eref{eq2}, \eref{fgrel2} and \eref{fgrel3}, we get
\begin{eqnarray}
\label{eq6}
dx'&=\frac{\partial f}{\partial x}dx+\frac{\partial f}{\partial t}dt\nonumber\\
&=\left(\frac{\partial f}{\partial x}u+\frac{\partial f}{\partial t}\right)dt\nonumber\\
&=(u-v)\frac{\partial g}{\partial t}dt \nonumber\\
&=u'(\frac{\partial g}{\partial x}u+\frac{\partial g}{\partial t})dt.
\end{eqnarray}
From the last two lines of Eq. \eref{eq6},
we find that the function $g$ satisfies the equation
\begin{eqnarray}
\label{eq7}
uu'\frac{\partial g}{\partial x}+(u'-u+v)\frac{\partial g}{\partial t}=0.
\end{eqnarray}
If $u'=u-v$, then we get the addition of velocities in Newtonian mechanics and we can derive the Galileo transformation.
This is in conflict with the principle of the constancy of the speed of light,
so it can be excluded, i.e., $u'\neq u-v$.
Similar to the solution of Eq. \eref{ft}, Eq. \eref{eq7} can be written as
\begin{eqnarray}
\label{eq7a}
\frac{\partial g}{\partial x}=-\frac{1}{\tilde{u}}\frac{\partial g}{\partial t},
\end{eqnarray}
and the solution is
\begin{eqnarray}
\label{g}
g=g(x-\tilde{u} t),
\end{eqnarray}
where $\tilde{u}=uu'/(u'-u+v)$.
The function $g(q)$ is an arbitrary function of a single variable $q$.
Up to this step, we don't know whether the constant $\tilde{u}$ depends on the motion of the body,
so we leave it as an arbitrary constant.
For light rays, $u'=u=c$, we get $\tilde{u}=c^2/v$.
If $\tilde{u}$ is a constant which depends only on $v$,
then $\tilde{u}=c^2/v$ and the above relation implies the relativistic addition of velocities.

{\bf Theorem}: The general coordinate transformation between inertial frames must be linear transformation.

Combining Eqs. \eref{fgrel3} and \eref{eq7a}, we get
\begin{eqnarray}
\label{fgrel1a}
\frac{\partial f}{\partial x}=-\tilde{u}\frac{\partial g}{\partial x}.
\end{eqnarray}
Eqs. \eref{fgrel2} and \eref{fgrel1a} can also be written as
\begin{eqnarray}
\label{fgreleq}
\frac{d f(p)}{d p}=-\tilde{u}\frac{d g(q)}{d q}.
\end{eqnarray}
If $\tilde{u}=v$, then from Eq. \eref{fgreleq}, we get $x'=f=-vg=-vt'$,
this contradicts the principle of the constancy of the speed of light,
so $\tilde{u}\neq v$.
Since the left hand side of Eq. \eref{fgreleq} is a function of $p=x-vt$ and the right hand side is a function of $q=x-\tilde{u}t$,
so Eq. \eref{fgreleq} means that the equality is a constant, 
henceforth $f(x-vt)$ and $g(x-\tilde{u}t)$ are linear functions.
The statement can also be proved as follows.
Take the partial derivative with respect to $t$ on both sides of Eq. \eref{fgrel1a}, we get
\begin{eqnarray}
\label{fgrel1b}
\frac{\partial^2 f}{\partial x\partial t}=-\tilde{u}\frac{\partial^2 g}{\partial x\partial t}.
\end{eqnarray}
Take the partial derivative with respect to $x$ on both sides of Eq. \eref{fgrel2}, we get
\begin{eqnarray}
\label{fgrel1c}
\frac{\partial^2 f}{\partial x\partial t}=-v\frac{\partial^2 g}{\partial x\partial t}.
\end{eqnarray}
From Eqs. \eref{fgrel1b} and \eref{fgrel1c}, we get
\numparts
\begin{eqnarray}
\label{fgrel1d}
\frac{\partial^2 f}{\partial x\partial t}=-v\frac{d^2f(p)}{dp^2}=0,\\
\label{fgrel1f}
\frac{\partial^2 g}{\partial x\partial t}=-\tilde{u}\frac{d^2g(q)}{dq^2}=0.
\end{eqnarray}
\numparts
Therefore, we prove that both $f=A(v)(x-vt)$ and $g=B(v)(x-\tilde{u}t)$ are linear functions, so the transformations between inertial frames are linear.
Furthermore, from the relation \eref{fgrel1a} and the constant $\tilde{u}=c^2/v$,
we get $B(v)=-v A(v)/c^2$ and $g=A(v)(t-vx/c^2)$.

In conclusion, we use the two postulates of special relativity and the assumption of the homogeneity
and isotropy of spacetime to prove that
the general coordinate transformation between inertial frames must be linear transformation.
Note that the postulate of the constancy of the speed of light is also used for clock synchronization and the definition of time so that we can define the inertial coordinate system.

\ack{
The authors would like to thank Jianwei Cui and Jerrold Franklin for fruitful discussions and comments. This research was supported in part by the Natural Science Foundation of China under Grant Nos. 12175184 and 11875136.}

\end{document}